# Non-EPR pairs quantum channel for teleporting an arbitrary two-qubit state

Xin-Wei Zha[*], Hai-Yang Song

Department of Applied Mathematics and Physics, Xi'an Institute of Posts and Telecommunications,

Xi'an, 710061 Shaanxi, China

Recently, Yeo and Chua [Phys. Rev. Lett. **96**, 060502 (2006)] have given an explicit protocol for faithfully teleporting an arbitrary two-qubit state via a genuine four-qubit entangled state, which is not reducible to a pair of Bell states. Here, we present a "transformation operator" to give the criterion of for faithfully teleporting arbitrary two-qubit states. The theoretical explanations of some quantum channels are given by transformation operators. Furthermore, a new four-qubit entangled state quantum channel is presented.

Since the first create of quantum teleportation protocol by Bennett [1], research on quantum teleportation has been attracting much attention both theoretically and experimentally in recent years due to its important applications in quantum calculation and quantum communication. For example, already there have been several experimental implementations [2-4] of teleportation and several other schemes of quantum teleportation have been presented [5-11]. Up to now, almost all complete set measurements are Bell state measurement and the states of entangled channel are reducible to a pair of Bell states.

Actually, there are many forms of quantum channel, which are not reducible to a pair of Bell states can be employed for a perfect teleportation arbitrary two-qubit state. Recently, Yeo and Chua [12] present a protocol for faithfully teleporting an arbitrary two-qubit state via a genuine four-qubit entangled state. By construction, their four-partite state is not reducible to a pair of Bell states. Chen [13] generalize completely the results of Ref. [12], to teleporting an arbitrary N-qubit state via genuine N-qubit entanglement channels. In this Letter, we present a new quantum channel for faithfully teleporting arbitrary two-qubit states employing genuine four-qubit entangled state $|\varphi\rangle_{3456}$ [Eq. (2)] that four-partite state is also not reducible to a pair of Bell states. The key element of our approach is a reversible operator $\sigma_{56}^{ij}$ [Eq. (3)] that we call "transformation operator". Later, we analyze the four-party Greenberger-Home-Zeilinger (GHZ) [14] and W [15] states using our method.

We first consider the teleportation of an arbitrary two-qubit state $|\chi\rangle_{12}$ using four-qubit

---

entangled state $|\varphi\rangle_{3456}$. If the sender Alice has two particles 1,2 in an unknown state

$$|\chi\rangle_{12} = (x_0|00\rangle + x_1|01\rangle + x_2|10\rangle + x_3|11\rangle)_{12} \qquad (1)$$

where $x_0, x_1, x_2$ and $x_3$ are arbitrary complex numbers, and we assume wave function satisfy normalization conditions $\sum_{i=0}^{3}|x_i|^2 = 1$. The entanglement channel between Alice and Bob is a four-qubit entangled state $|\varphi\rangle_{3456}$. The particle pair (1,2) and particles 3, 4 are in Alice's possession, and other two particles 5, 6 are in Bob's possession. The system state of six particles can be expressed as

$$|\psi\rangle_{123456} = |\chi\rangle_{12} \otimes |\varphi\rangle_{3456} \qquad (2)$$

In accordance with the principle of superposition and transformation operator, $|\psi\rangle_{123456}$ can be represented in the form of a series [16]

$$|\psi\rangle_{123456} = |\chi\rangle_{12} \otimes |\varphi\rangle_{3456} = \frac{1}{4}\sum_{i=1}^{4}\sum_{j=1}^{4} \varphi_{13}^{i} \varphi_{24}^{j} \sigma_{56}^{ij} |\chi\rangle_{56} \qquad (3)$$

where $\varphi_{13}^{i}, \varphi_{24}^{j}$ are the Bell states, and

$$\varphi_{mn}^{1} = \frac{1}{\sqrt{2}}(|00\rangle + |11\rangle)_{mn},$$

$$\varphi_{mn}^{2} = \frac{1}{\sqrt{2}}(|00\rangle - |11\rangle)_{mn},$$

$$\varphi_{mn}^{3} = \frac{1}{\sqrt{2}}(|01\rangle + |10\rangle)_{mn},$$

$$\varphi_{mn}^{4} = \frac{1}{\sqrt{2}}(|01\rangle + |10\rangle)_{mn} \qquad mn = 13, 24 \qquad (4)$$

$$|\chi\rangle_{56} = (x_0|00\rangle + x_1|01\rangle + x_2|10\rangle + x_3|11\rangle)_{56} \qquad (5)$$

The operators $\sigma_{56}^{ij}$ are called the "transformation operators". If the $\sigma_{56}^{ij}$ are unitary operators, then Alice informs Bob two Bell state measurement outcomes via a classical channel. By outcomes received, Bob can determine the state of particles 5,6 exactly by transformation operators $(\sigma_{56}^{ij})^{-1}$. The unknown two-particle entangled state can be teleported perfectly, and the successful possibilities and the fidelities of the two schemes both reach unit. If the $\sigma_{56}^{ij}$ are not unitary operators, the determinant of transformation operators are not zero. As ref [16], Bob introduces an auxiliary two-state particle $a$ with the initial state $|0\rangle_a$ and performs a collective unitary transformation on particles 5, 6 and *a*. Then Bob measures the state of particle *a*. If the measured result is $|1\rangle_a$, the teleportation is failed. If the measurement result is $|0\rangle_a$, the teleportation is successfully realized. The probability of successful teleportation is small unit. If the determinants of transformation operators are zero, the unknown two-particle arbitrary entangled state can't be teleported perfectly.

Also we can write

$$|\psi\rangle_{123456} = |\chi\rangle_{12} \otimes |\varphi\rangle_{3456} = \frac{1}{4}\sum_{i,j=1}^{4} |g^{ij}\rangle_{1234} \sigma_{56}^{ij} |\chi\rangle_{56} \quad (3')$$

where $|g^{ij}\rangle = \varphi_{13}^{i}\varphi_{24}^{j}$ are the 16 G states by Rigolin [7].

As an example, we now consider the quantum channel in Ref. [12]

$$|\varphi\rangle_{3456} = \frac{1}{2\sqrt{2}}(|0000\rangle - |0011\rangle - |0101\rangle + |0110\rangle + |1001\rangle + |1010\rangle + |1100\rangle + |1111\rangle)_{3456} \quad (6)$$

It is easy to obtain transformation operator

$$\sigma_{56}^{11} = \begin{pmatrix} \frac{1}{\sqrt{2}} & 0 & 0 & \frac{1}{\sqrt{2}} \\ 0 & \frac{1}{\sqrt{2}} & -\frac{1}{\sqrt{2}} & 0 \\ 0 & \frac{1}{\sqrt{2}} & \frac{1}{\sqrt{2}} & 0 \\ -\frac{1}{\sqrt{2}} & 0 & 0 & \frac{1}{\sqrt{2}} \end{pmatrix} \quad (7)$$

Furthermore, we can obtain other transformation operators

$$\hat{\sigma}_{56}^{ij} = \hat{\sigma}_{56}^{11}\left(\sigma_5^i \otimes \sigma_6^j\right) \quad (8)$$

where $\hat{\sigma}_m^k = I_m, \sigma_{mz}, \sigma_{mx}, -i\sigma_{my}$. $m = 5,6$, $I_m$ is the two-dimensional identity and $\sigma_{mz}, \sigma_{mx}, \sigma_{my}$ are the Pauli matrices.

Apparently, if the $\hat{\sigma}_{56}^{11}$ is unitary operator, the $\hat{\sigma}_{56}^{ij}$ are also unitary operators. Therefore, the teleportation can be realized only by performing an inverse transformation.

If the quantum channel is GHZ state, i.e. $|\varphi\rangle_{3456} = \frac{1}{2}(|0000\rangle + |1111\rangle)_{3456}$

Then the transformation operator $\sigma_{56}^{11} = \begin{pmatrix} \sqrt{2} & 0 & 0 & 0 \\ 0 & 0 & 0 & 0 \\ 0 & 0 & 0 & 0 \\ 0 & 0 & 0 & \sqrt{2} \end{pmatrix}$; $\quad (9)$

If the quantum channel is W state, $|\varphi\rangle_{3456} = \frac{1}{2}(|0001\rangle + |0010\rangle + |0100\rangle + |1000\rangle)_{3456}$

Then
$$\sigma_{56}^{11} = \begin{pmatrix} 0 & 1 & 1 & 0 \\ 1 & 0 & 0 & 0 \\ 1 & 0 & 0 & 0 \\ 0 & 0 & 0 & 0 \end{pmatrix}. \tag{10}$$

From Eqs. (9) and (10), it is easy to see that the determinants of transformation operators are zero. Therefore, we can't use the GHZ states and the *W* states [7] to deterministically teleport arbitrary two qubits.

Now, let us assume Alice and Bob share a entangled channel

$$|\varphi\rangle_{3456} = \frac{1}{2}(|0000\rangle + |0101\rangle + |1011\rangle + |1110\rangle)_{3456} \tag{11}$$

Using Eqs. (3), we can obtain the transformation operator

$$\sigma_{56}^{11} = \begin{pmatrix} 1 & 0 & 0 & 0 \\ 0 & 1 & 0 & 0 \\ 0 & 0 & 0 & 1 \\ 0 & 0 & 1 & 0 \end{pmatrix} \tag{12}$$

Obviously, the $\sigma_{56}^{11}$ is a unitary operator and the $\sigma_{56}^{11}$ is also a C-not transformation operator. Therefore, Bob can determine the unknown two-particle entangled states by performing C-not operation with the particle 5 as the control qubit and the particle 6 as the target.

Recently, Chen [13] consider the teleportation of the same *N* qubit state using the maximally entangled state（MES）channel between Alice and Bob. Zhan-jun Zhang [17] present the quantum channel linking Alice and Bob is the state $|Q_{ij}\rangle_{A_1B_1A_2B_2}$ and it can be equivalently written in the form of $U_{A_1A_2}|\psi_i\rangle_{A_1B_1}|\psi_j\rangle_{A_2B_2}$. It is easy to show that the transformation operators are unitary operators for those quantum channels. Therefore, the teleportation can be also realized only by performing an inverse transformation.

In conclusion, we have presented a new method for faithfully teleport an arbitrary two-qubit state employing genuine four-qubit entangled states $|\varphi\rangle_{3456}$ that four-partite state is also not reducible to a pair of Bell states. The key element of our approach is a reversible operator $\sigma_{56}^{ij}$ [Eq. (3)] that we call "transformation operators". The "transformation operators" give the criterion of for faithfully teleporting arbitrary two-qubit states. As examples, we analyze the quantum channel in Ref. [13] and the four-party Greenberger-Home-Zeilinger (GHZ) [14] and W [15] states using our method. The relation of transformation operator with the entanglement of quantum channel and orthogonal joint measurement will be further investigated.

## ACKNOWLEDGEMENT

This work is supported by Shaanxi Natural Science Foundation under Contract Nos. 2004A15 and Science Plan Foundation of office the Education Department of Shaanxi Province Contract Nos. 05JK288.

## REFERENCES

[1] C.H. Bennett, G. Brassard, C. Crepeau, et al., Phys. Rev. Lett. **70**, 1895 (1993).

[2] D. Bouwmeester, J.W. Pan, K. Mattle, et al., Nature (London) **390**, 575 (1997).

[3] J.W. Pan, M. Daniell, S. Gasparoni, et al., Phys. Rev. Lett. **86**, 4435 (2001).

[4] Z. Zhao, Y.A. Chen, A.N. Zhang, et al., Nature (London) **430**, 54 (2004).

[5] W.L. Li, C.F. Li, G.C. Guo, Phys. Rev. A **61**, 034301 (2000).

[6] H. Lu, G.C. Guo, Phys. Lett. A. **276**, 209. (2000).

[7] G. Rigolin, Phys. Rev. A **71**, 032303 (2005).

[8] F.L. Yan, H.W. Ding, Chin. Phys. Lett. **23**, 17 (2006).

[9] J.X. Fang, Y.S. Lin, S.Q. Zhu, et al., Phys. Rev. A **67**, 014305 (2003).

[10] L. Roa, A. Delgado, I. Fuentes-Guridi, Phys. Rev. A. **68**, 022310 (2003).

[11] G. Gordon, G. Rigolin, Phys. Rev. A. **73**, 042309 (2006).

[12] Y. Yeo, W.K. Chua, Phys. Rev. Lett. **96**, 060502 (2006).

[13] P.X. Chen, S.Y. Zhu, G.C. Guo, Phys. Rev. A **74**, 032324 (2006).

[14] D.M. Greenberger, M.A. Horne, A. Zeilinger, in *Bell's Theorem, Quantum Theory, and Conceptions of the Universe*, edited by M. Kafatos (Kluwer Academic,Dordrecht, 1989), pp. 69–72.

[15] A. Zeilinger, M.A. Horne, D.M. Greenberger, in *Proceedings of Squeezed States and Quantum Uncertainty*, edited by D. Han, Y.S. Kim, and W.W. Zachary, NASA Conference Publication No. 3135(NASA, Washington, DC, 1992), pp. 73–81.

[16] X.W. Zha, C.B. Huang, e-print quant-ph/.0607116.

[17] Z. J. Zhang, e-print quant-ph/.0611290.